\documentclass[a4paper, mathleft]{an}

\usepackage{times}
\usepackage{graphicx}
\usepackage{natbib}
\usepackage{nicefrac}
\usepackage{amsmath}
\usepackage{amssymb}
\usepackage[usenames, dvipsnames]{color}
\usepackage{multirow}
\usepackage[rightcaption]{sidecap}
\usepackage{units}

\usepackage{hyperref}
\hypersetup{
    final=true,
    pageanchor=true,
    colorlinks=true,
    breaklinks=true,
    linkcolor=blue,
    citecolor=blue,
    urlcolor=blue,
    pdfpagemode=UseNone}

\newcommand\phn{\phantom{0}}
\newcommand\phm{\phantom{$-$}}

\bibpunct{(}{)}{;}{a}{}{,}
\definecolor{myGreen}{rgb}{.2,.8,.2}

\newcommand{\arcdeg}{\mbox{$^{\circ}$}}

\begin{document}

\Pagespan{1}{}
\Yearpublication{2015}
\Yearsubmission{2015}
\Month{1}
\Volume{336}
\Issue{1}
\DOI{10.1002/asna.2015xxxxx}

\title{Digitization of sunspot drawings by Sp\"orer made in 1861--1894}
\author{
    A.\ Diercke\inst{1,2}\fnmsep\thanks{Corresponding author:
    {adiercke@aip.de}}
\and
    R.\ Arlt\inst{1}
\and
    C.\ Denker\inst{1}}
\titlerunning{Digitization of sunspot drawings by Sp\"orer 1861--1894}
\authorrunning{A. Diercke, R. Arlt \& C. Denker}
\institute{
    Leibniz-Institut f\"ur Astrophysik Potsdam,
    An der Sternwarte 16,
    14482 Potsdam, Germany
\and
    Universit\"at Potsdam,
    Institut f\"ur Physik und Astronomie,
    Karl-Liebknecht-Stra\ss{}e 24/25,
    14476 Potsdam,
    Germany 
    }
\received{October 2, 2014}
\accepted{November 17, 2014}
\publonline{January 2015}


\abstract{Most of our knowledge about the Sun's activity cycle arises from 
sunspot observations over the last centuries since telescopes have been used for 
astronomy. The German astronomer Gustav Sp\"orer observed almost daily the Sun 
from 1861 until the beginning of 1894 and assembled a 33-year collection of 
sunspot data covering a total of 445 solar rotation periods. These sunspot 
drawings were carefully placed on an equidistant grid of heliographic longitude 
and latitude for each rotation period, which were then copied to copper plates 
for a lithographic reproduction of the drawings in astronomical journals. In 
this article, we describe in detail the process of capturing these data as 
digital images, correcting for various effects of the aging print materials, and 
preparing the data for contemporary scientific analysis based on advanced image 
processing techniques. With the processed data we create a butterfly diagram 
aggregating sunspot areas, and we present methods to measure the size of 
sunspots (umbra and penumbra) and to determine tilt angles of active regions. A 
probability density function of the sunspot area is computed, which conforms to 
contemporary data after rescaling.}
\maketitle

\section{Introduction}
With the advent of the telescope 400 years ago the longest record of direct
evidence of solar activity  began. Sunspots are emergences of localized magnetic
flux, but their statistics indicate that they are also manifestations of a
large-scale magnetic field in the solar interior. There are many statistical
properties that can be derived from sunspots such as their spatial distributions
across the hemispheres as well as within sunspot groups, their areas and
life-times, or their rotation and motion relative to the photosphere. It is
desirable to construct time series of sunspot positions and areas of as much of
the 400-year period as possible.

Long series of observations in the pre-photographic period are available from, 
e.g., Johann Caspar Staudacher who observed in 1749--1796
\citep{Arlt2008, Arlt2009a}, James Archibald Hamilton and William Gimingham 
1795--1797 \citep{Arlt2009b}, Honor\'{e} Flaugergues 1788--1830 
\citep{Wolf1861}, Samuel Heinrich Schwabe 1825--1867 \citep{Arlt2013}, Richard 
Christopher Carrington  1853--1861 \citep{Carrington1863}, and Gustav Sp\"orer 
1861--1894 (present work). The references point to detailed analyses, and some 
of them include positions. Meanwhile the first regular photographic observations 
started in Greenwich in 1874 \citep{Baumann2005}; data of earlier visual 
observations are also available. 

Due to the nature of seeing, the photographs of the Greenwich observations show 
much less detail than visual data, rendering the early Greenwich data inferior. 
In the subsequent analysis of Sp\"orer's sunspot data, only the average 
positions of groups and total areas were compiled in a database, while 
additional, useful information can be obtained from individual sunspot positions 
and areas which are contained in the visual data set. If the visual data is 
recovered and analyzed, it becomes possible to estimate the area of individual 
sunspots and to determine the tilt angles of bipolar groups as well as other 
properties depending on the progress of the solar cycle.

The original visual observations by Sp\"orer are apparently lost, while his 
printed articles contain detailed synoptic maps of the Sun with the observed 
sunspot groups. They contain much more information than Sp\"orer's own 
measurements.

Sp\"orer's measurements from 1883--1893, collected in tables, were used to
calculate the solar rotation and compare it with the solar rotation
derived from observations made in Greenwich and Kanzelh\"ohe
\citep{Balthasar1988, Woehl1989}.

The present study aims at adding one mosaic piece to the 400-year record of
sunspot observations by extracting the positional and area information from the
visual observations by Gustav Sp\"orer who observed in 1861--1894.

\begin{figure}[t]
\centering
\includegraphics[width=\columnwidth]{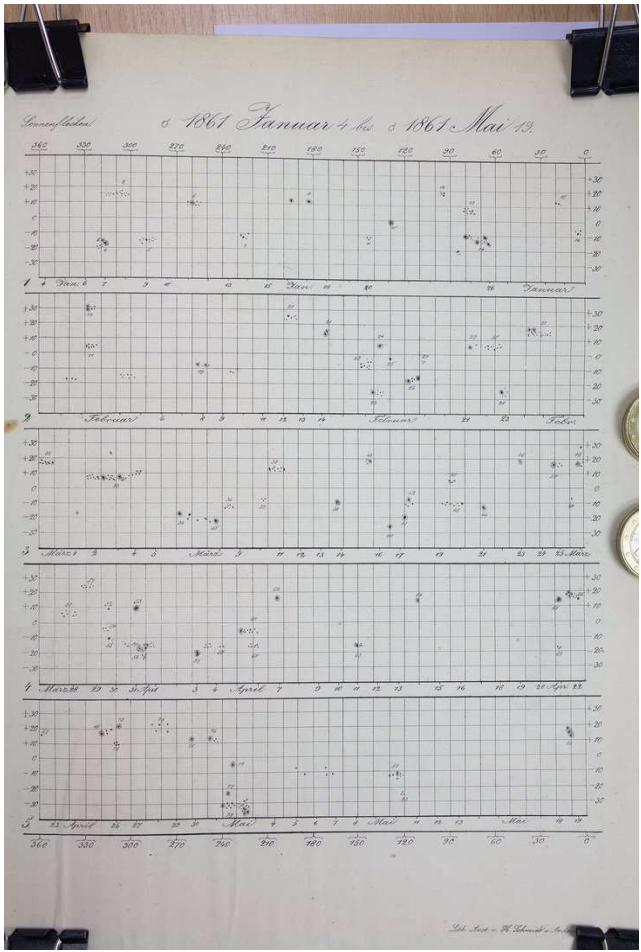}
\caption{Original drawings with five solar rotation periods after digitization. 
The title line shows the period of observations. The number of each rotation 
period and the observing date are marked in the left corner and under each 
rotation period, respectively.  Binder clips and heavy metal coins are used to 
flatten the paper.}
\label{fig:orgim}
\end{figure}

\section{Observations}
Born in 1822 in Berlin, Friedrich Wilhelm Gustav Sp\"orer studied mathematics 
and astronomy at the University of Berlin. In 1843 Sp\"orer obtained his degree 
and started to work on his dissertation at the Observatory in Berlin. After 
becoming an instructor in Anklam in 1849 his interest in sunspots peaked and led 
to the start of his observations on 2 December 1860 \citep{Spoerer1861}. He 
published his results in numerous articles in \textit{Publicationen des 
Astrophysikalischen Observatoriums zu Potsdam} and \textit{Publicationen der 
Astronomischen Gesellschaft} \citep{Spoerer1874, Spoerer1878, Spoerer1880, 
Spoerer1886, Spoerer1894}. Later, in 1874, he moved to Potsdam and worked at the 
Astrophysical Observatory, where he continued his observations of sunspots until 
1894. Sp\"orer died in 1895 in Gie\ss{}en while traveling \citep{Vogel1895}.

Sp\"orer is famous for his simplification of  Richard Carrington's theory about 
sunspot latitudes in a solar cycle which is known as \textit{Sp\"orer's Law}. It 
was also Sp\"orer who found a minimum of solar activity between about 1645 and 
1715 with the few spots appearing between 1672 and 1704 all being on the 
southern hemisphere \citep{Spoerer1887}. This minimum (1645--1715) was named 
after the English astronomer Edward Walter Maunder. But an earlier minimum from 
1420--1570, found with the $^{14}$C method \citep{Usoskin2007, Bard1997}, is now 
termed Sp\"orer minimum in honor of Sp\"orer's discovery of the Maunder minimum 
\citep{Eddy1976}.

Long continuous observations of sunspots were made by Samuel Heinrich Schwabe
1825--1867 \citep{Arlt2011a}. In this time period,  Carrington started his
observations in 1853, which were continued by Sp\"orer's observations beginning
in January 1861 during Carrington rotation period No.~96. Carrington arranged
the sunspots of one solar rotation period according to their heliographic
longitudes and latitudes.  Sp\"orer used the same method allowing to connect
both data sets. There are two ways in which Sp\"orer provides us with data on
sunspots:
  \begin{itemize}
  \item Sunspot synoptic maps similar to Carrington's synoptic charts but differ
        by an offset in longitude.
  \item Tables with measurements of sunspot positions on various days of the
        visibility period of a group. In this way, the disk passage of the
        sunspot groups can be visualized.
 \end{itemize}

Sp\"orer used different telescopes for his observations. First he used a
3\nicefrac{1}{2}-foot telescope and after 1865 a 7-foot telescope
\citep{Spoerer1874}. Units commonly used in Prussia at that time are given in
\citet{Noback1851}. In 1868, he received a new telescope, with a 5-inch
Steinheil lens, as a present from the crown prince of Prussia for his research
\citep{Vogel1895, Spoerer1874}. Since 1879 he observed with a
7\nicefrac{1}{2}-inch diameter Grubb refractor \citep{Spoerer1878, Spoerer1880,
Spoerer1886, Spoerer1894} at the Astrophysical Observatory Potsdam. After 1884
photographic plates were used for the observations. But Sp\"orer still observed
sunspots visually with the telescope. In general, it was easier to find sunspots
first with the telescope before capturing them on photographic plates. The
telescope was also helpful when the sky was cloudy. During short breaks in the
clouds it was possible to quickly gauge sunspot positions and shapes visually,
while it was hopeless to take a photograph. In bad seeing conditions bigger
sunspots were still faintly visible on the photographic plates but small spots
were indeterminable \citep{Spoerer1894}.

\begin{figure*}[t]
\centering
\includegraphics[width=\columnwidth]{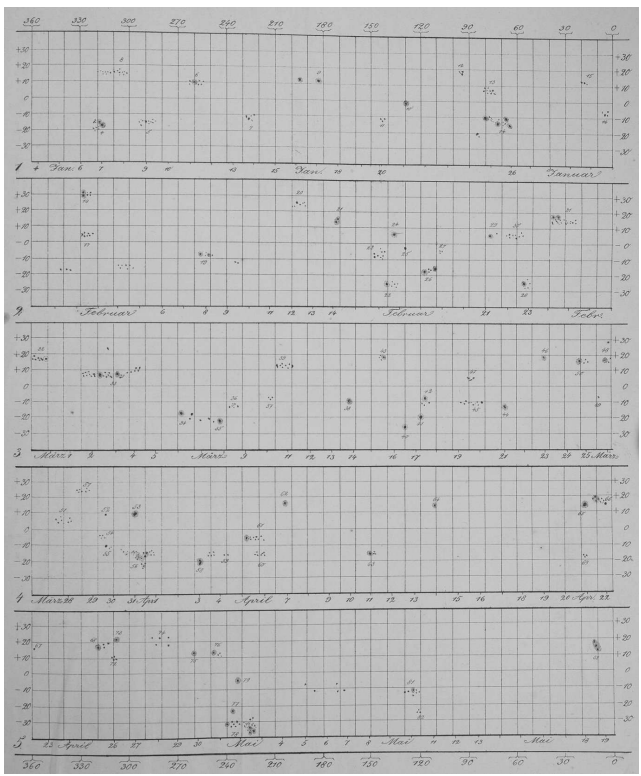}
\hfill
\includegraphics[width=1\columnwidth]{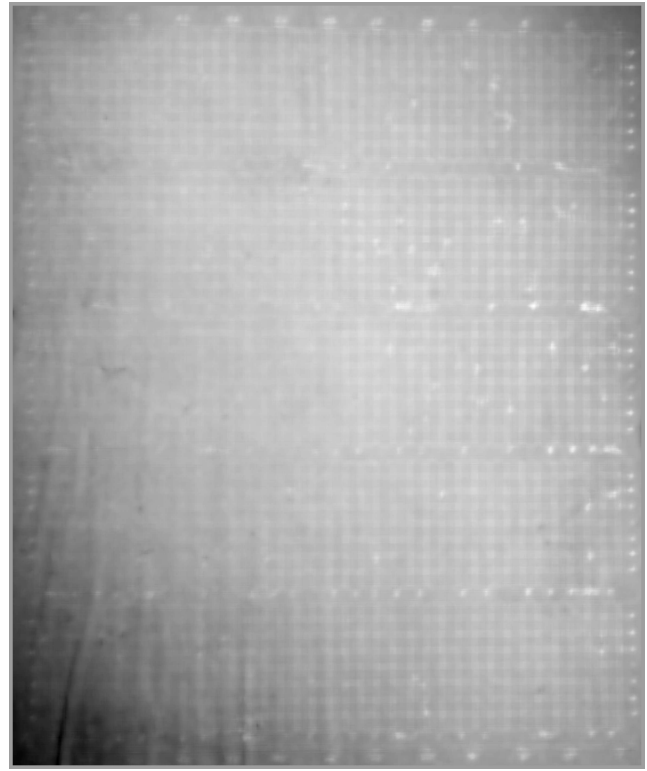}
\caption{Rotated and cropped image (\textit{left}) after correction by a 
background image (``pseudo flat-field'') derived from an order-statistics
filter 
(\textit{right}). Both images are displayed on a linear scale between the 
minimum and maximum of their respective intensities.} 
\label{FIG}
\end{figure*}

If observations were out of question because of bad weather or other
circumstances, Sp\"orer used the data from other observatories in Vienna,
M\"unster, or Rome to augment his data set \citep{Spoerer1874}. Auxiliary
entries are marked with characters in the data tables. Especially in the
beginning he received data from Prof. Eduard Heis in M\"unster, which he labeled
with an $H$. Sp\"orer carried out most of his observations at local noon. He
noted date and time of his sunspot observations, and he measured the position
angle $p$ relative to the terrestrial equator and the geocentric angular
distance $\rho$ from the middle of the solar disk. With these two variables it
is possible to calculate the heliographic coordinates, i.e., the longitude $l$
and the latitude $b$. Once longitude and latitude are known, the normalized
longitude $L$ as shown in Fig.~\ref{fig:orgim} can be computed \citep[for
details see][]{Spoerer1874}. The coordinates $b$ and $L$ of the day of meridian
passage were used to draw the sunspots. The abscissa shows the normalized
longitude $L$ from 360$^{\circ}$ to 0$^{\circ}$ while the ordinate is the
heliographic latitude $b$ in the range of $\pm 40^{\circ}$ for one rotation
period. Each of the 89 pages contains five rotation periods (Fig.
\ref{fig:orgim}) amounting to 445 rotation periods in total. The sunspot
drawings were published as lithographs, which even show details of sunspots such
as umbra and penumbra (Fig.~\ref{fig:spotgroup}).

\section{Image processing}

\subsection{Image capture and digitization}

The sunspot drawings were collated in hardbound books with rigid covers, where 
several signatures (folded sheets of paper) are stitched together in the books' 
spines. Consequently, pages will not be perfectly flat when opening the books. 
Binder clips and heavy metal coins are used to flatten the drawings 
(Fig.~\ref{fig:orgim}) in preparation for acquiring digital images. All images 
are taken by artificial light from two bright desk lamps under the same 
conditions.

Whole pages are photographed with a Canon EOS 5D digital camera, which records 
images with $4368\times2912 $ pixels. The Sigma 50~mm $f / 2.8$ EX DG macro lens 
ensures that image distortions are minimal over the field-of-view, while small 
perspective errors may remain and are easily corrected later. The image scale of 
the recorded images is about 0.35~mm pixel$^{-1}$. The 36-bit digital images are 
preprocessed with the computer program 
RawTherapee\footnote{\href{http://rawtherapee.com/}{rawtherapee.com}} 
Version~2.4.1 and saved as 24-bit RGB images in the Tagged Image File Format 
(TIFF). The filenames (\texttt{spoererYYYYMMDD$\_$YYYYMMDD.tif}) are created 
using the start and end dates of the first and last rotation periods depicted in 
the images, respectively.

\subsection{Removal of artifacts and conditioning of images}

\begin{figure*}[t]
\centering
\includegraphics[width=\textwidth]{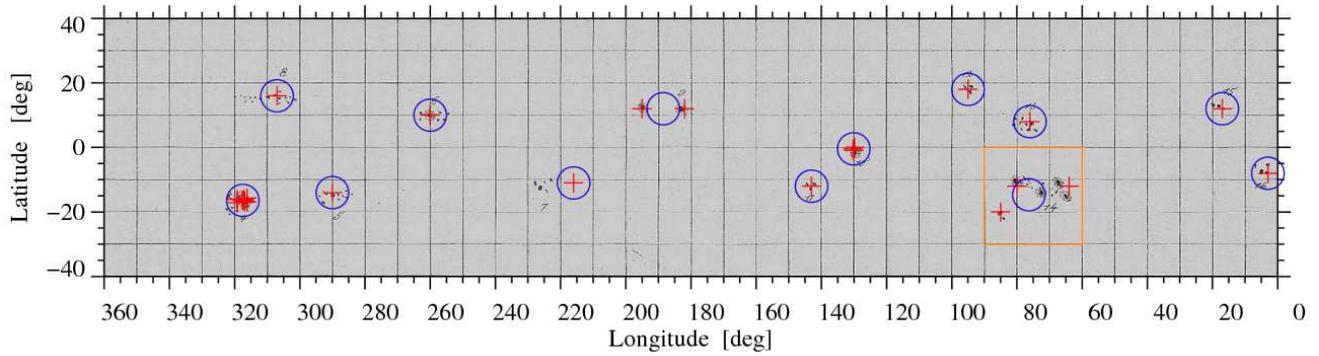}
\caption{Image after remapping with identified regions from Sp\"orer's
measurements. The red crosses refer to the positions of the sunspots and the
blue circles mark the mean positions of the sunspot groups. The positions were
measured over several days, i.e., it is possible to visualize the transit of the
sunspot group across the solar disk. Sometimes there are errors in the data (see
Sect.~\ref{sec:errors}), e.g., the table entry of spot group No.~7 differs from
its position in the longitude-latitude map. Properties of spot group No.~14
(\textit{orange square}) are discussed in Sect.~\ref{sec:position}. }
\label{Fig:map}
\end{figure*}

The sunspot drawings were published nearly 150 years ago, thus, the pages are
yellowish, stained, and corrugated. These artifacts have to be removed because
they will severely affect the analysis of sunspot properties. All algorithms
hereafter are developed in the Interactive Data Language
(IDL)\footnote{\href{http://www.exelisvis.com/}{www.exelisvis.com}}. For the 
most part, the programs run with a minimum of user interaction to avoid 
potential biases.

Initially, the three-component color images are converted to gray-scale images 
by simply averaging their color planes. The images are rotated by $90\arcdeg$ to 
have them in portrait format as in the books. The bitmaps are then cropped so 
that only the coordinate grids with sunspot data and their annotations remain. 
In the next processing step,  a small position angle error of the camera has to 
be measured and removed. The grid lines are the dominant feature in the images, 
thus, averaging all rows and all columns results in two intensity traces with 
sharp minima at the $x$- and $y$-coordinates of the grid lines, respectively. 
The standard deviation of the intensity traces will reach a maximum, when the 
angular offset is zero. Therefore, we rotate the images in steps of $0.1\arcdeg$ 
within the range of $\pm 1\arcdeg$ and compute the respective standard 
deviations. A parabolic fit to this curve yields the exact angle by which the 
images have to be rotated. Cubic spline interpolation \citep{Park1983} is used 
to minimize interpolation errors between pixels. An order-statistics filter is 
used to compute a background intensity map, thus, eliminating the grid lines, 
annotations, and sunspots (Fig.~\ref{FIG}). The most common order-statistics 
filter are minimum, maximum, and median filters \citep[e.g.,][]{Gonzalez2002}. 
However, none of them provides an acceptable map of background intensities. 
Therefore, we implemented a filter,  which ranks the intensities within a 
33$\times$33-pixel neighborhood and replaces the value of the center pixel 
with the value that represents the $80^\mathrm{th}$ percentile of the intensity 
distribution. The 16-pixel wide borders of the full background intensity map are 
simply replaced by the mean background intensity.

The effect of such a non-linear spatial filter is to replace clusters or lines 
of dark pixels by gray levels, which resemble more closely their neighborhood. 
Closer inspection of the right panel in Fig.~\ref{FIG} reveals that the 
procedure captures well the global intensity trends, which are evident in the 
lower-left and upper-right corners. In addition, faint shadows created by the 
wrinkled paper are efficiently removed. What appears at a first glance as a 
potentially bothersome artifact, i.e., a faint brightening associated with grid 
lines and annotations, augments the contrast in the next image processing step. 
Division of an image by its background intensity map, which correspond to a 
pseudo flat-fielding procedure, results in an artificially ``bleached'' and 
flattened image (left panel in Fig.~\ref{FIG}). This corrected image still 
contains distortions due to wrinkling of the paper and because of the curvature 
of pages photographed from an opened hardbound book.

\begin{figure*}[t]
\centering
\includegraphics[width=\textwidth]{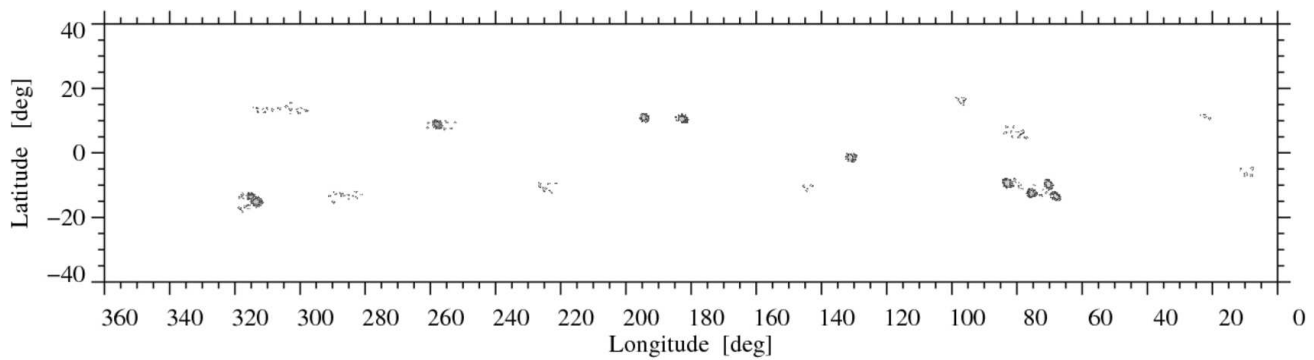}
\caption{Image after applying morphological image processing to remove the grid 
lines. The sunspots are now clearly visible. In addition, the sunspot group 
numbers have been removed with a point-and-click procedure.}
\label{Fig:test2}
\end{figure*}

\subsection{Geometric correction and resampling of images}

The first step in geometrically correcting distorted images is to extract
the exact pixel coordinates, where the horizontal and vertical grid lines
intersect. We construct a 17$\times$17-pixel kernel by setting the values of
two intersecting, three-pixel wide horizontal and vertical lines to unity and
keeping all other pixels at zero. Convolving the corrected image with such a
kernel enhances the grid lines, in particular, where the lines cross each other.
Simple thresholding then yields a binary mask in which pixels belonging to the
grid have values of unity. Initial estimates of the grid coordinates can be
computed from the average profiles of rows and columns, which are smoothed by a
17-pixel wide filter \citep{Lee1986} and subsequently normalized. Parabolic fits
to the local maxima of the row and column profiles are in most cases good
estimates of the $x$- and $y$-grid coordinates. Note that the first 51 images
contain an extra axis at the bottom of the page displaying the heliographic
longitude $L$ in increments of 30\arcdeg. Usually, it is more difficult to
determine the  $y$-coordinates because of the preferential warping of the pages.
Therefore, the average column profiles are computed locally, i.e., only for
regions between adjacent vertical grid lines.

In order to determine the maximum in $x$- and $y$-direction, we take the grid
line between two grid points and determine the maximum of intensity of this
region. Direct-neighbor coordinates are used for points on the border.

Some of the grid coordinates are slightly misplaced. This type of error occurs
especially near sunspots. To remedy this, the mean distance to direct neighbors
of a grid point is estimated, i.e., the difference between mean distance and
actual distance to neighbors can be calculated. This difference should be less
than six pixels, otherwise the distance of the neighbors will be replaced by the
mean distance. In rare cases (only 10 occurrences) the  grid coordinates are off
by 30 pixels. In these special cases, the coordinates are replaced by the mean
coordinates of columns or rows in this rotation period.

In order to improve the  preliminary $x$- and $y$-coordinates, we extract  an
8$\times$8-pixel square region from a binary mask, enlarge this region by a
factor of 20 using cubic spline interpolation  and determine the intensity
maximum of this region. 

During the second to last step, we take a  140$\times$140-pixel region and
double its size. Then the grid is extracted by using a mask to minimize the
influence of a sunspot by estimating the maximum intensity of the grid line. The
highest intensity appears where two grid lines intersect. The intensity map  is
smoothed twice  using a 141$\times$141-pixel boxcar average to eliminate some
spurious features obfuscating the maximum. The maximum in $x$- and $y$-direction
is then the new grid coordinate. 

Finally, we visually examined all grid coordinates and replace manually the
remaining ones that do not match. For this purpose, the region is enlarged in a
separate window, and a better coordinate can be chosen by a point-and-click
operation with the computer mouse. On average we replaced 1.4 grid points for
each rotation period (37$\times$9 grid points).

The heliographic longitudes and latitudes are given in increments of
$10\arcdeg$, so that we have a grid of (37$\times$9) $x$- and $y$-coordinates
for each rotation period, which is sufficient to resample the distorted grid to
a new, equidistantly sampled grid with a resolution of 0.1\arcdeg\ pixel$^{-1}$.
We also include a 5\arcdeg-wide border to minimize sampling errors at the
periphery. Subsequently, Delaunay triangulations \citep{Lee1980} are used to
interpolate the sunspot data to the new grid (Fig.~\ref{Fig:map}). Data of this
type ($3600\times800$ pixels) are obtained for each of the 445 rotation periods
recorded by Sp\"orer and build the input for any further processing.

\subsection{Error sources} \label{sec:errors}

An electronic version of Sp\"orer's tabulated measurements was provided by Ilkka
Tuominen, Helsinki, which contains the sunspot positions from 1861 to 1884. Note
that this period is shorter than the time interval covered by the synoptic maps.
Using this table, we compare the sunspot positions in the drawings and the
coordinates in the tables (see Fig.~\ref{Fig:map}). In most cases, the
coordinates  are identical, except for a few cases, e.g., spot No.~7 in
Fig.~\ref{Fig:map}, where the positions differ. Table~\ref{tab:01} provides the
differences in latitude or longitude within the first 100~rotation periods. Most
of the erroneous coordinates deviate by exactly 10$^{\circ}$, but there are also
sign errors. We found 18~mismatches within 100~rotation periods with 984
sunspots.

\begin{table}[t]
\caption{Differences in sunspot positions between Sp\"orer's measurements and
coordinates in the first 100 synoptic maps.}
\footnotesize
\begin{tabular}{ccccc}
\hline\hline
\rule[0mm]{0mm}{3.5mm} Period & Sunspot & Longitude $L$ & Coordinates  &
Coordinates \\
\rule[-1.5mm]{0mm}{3mm}       & number  & or latitude $b$ & in the table & in
the image \\
\hline
 \phn 1 & \phn\phn 7 & $L$ &       \phm 216$^{\circ}$ & \phm
226$^{\circ}$\rule[0mm]{0mm}{3.5mm} \\
 \phn 3 &    \phn 46 & $L$ &    \phm\phn 31$^{\circ}$ & \phm\phn 41$^{\circ}$\\
 \phn 9 &        128 & $L$ &       \phm 228$^{\circ}$ & \phm 238$^{\circ}$\\
 \phn 9 &        130 & $L$ &       \phm 171$^{\circ}$ & \phm 181$^{\circ}$\\
     16 &    \phn 21 & $L$ &       \phm 109$^{\circ}$ & \phm 119$^{\circ}$\\
     33 &    \phn 69 & $b$ &    \phm\phn 27$^{\circ}$ & \phm\phn 17$^{\circ}$\\
     33 &    \phn 76 & $b$ &       \phn $-18^{\circ}$ & \phm\phn 18$^{\circ}$\\
     35 &    \phn 93 & $L$ &       \phm 304$^{\circ}$ & \phm 314$^{\circ}$\\
     39 &        132 & $L$ &       \phm 128$^{\circ}$ & \phm 138$^{\circ}$\\
     50 &        113 & $b$ &    \phm\phn 12$^{\circ}$ & \phm\phn 22$^{\circ}$\\
     53 &        140 & $L$ &       \phm 254$^{\circ}$ & \phm 264$^{\circ}$\\
     55 & \phn\phn 5 & $b$ &       \phn $-11^{\circ}$ & \phn $-15^{\circ}$\\
     62 &    \phn 89 & $b$ & \phm\phn\phn 9$^{\circ}$ & \phm\phn 19$^{\circ}$\\
     63 &        103 & $L$ &    \phm\phn 53$^{\circ}$ & \phm\phn 43$^{\circ}$\\
     68 &        138 & $L$ &       \phm 307$^{\circ}$ & \phm 302$^{\circ}$\\
     69 &    \phn 14 & $L$ &       \phm 218$^{\circ}$ & \phm 207$^{\circ}$ \\
     79 &    \phn 67 & $L$ &    \phm\phn 43$^{\circ}$ & \phm 102$^{\circ}$\\
     93 &    \phn 41 & $b$ & \phm\phn\phn 8$^{\circ}$ & \phm\phn
     18$^{\circ}$\rule[-1.5mm]{0mm}{3mm}\\
\hline
\end{tabular}
\begin{center}
\parbox{0.85\columnwidth}{\vspace*{-4mm}
\begin{itemize}
\item[Note:] Most positions have differences of exactly 10$^{\circ}$. We
  expect them to be mistakes in the lithographic prints. 
\end{itemize}}
\end{center}
\label{tab:01}
\end{table}

\begin{figure*}[t]
\centering
\includegraphics[width=\textwidth]{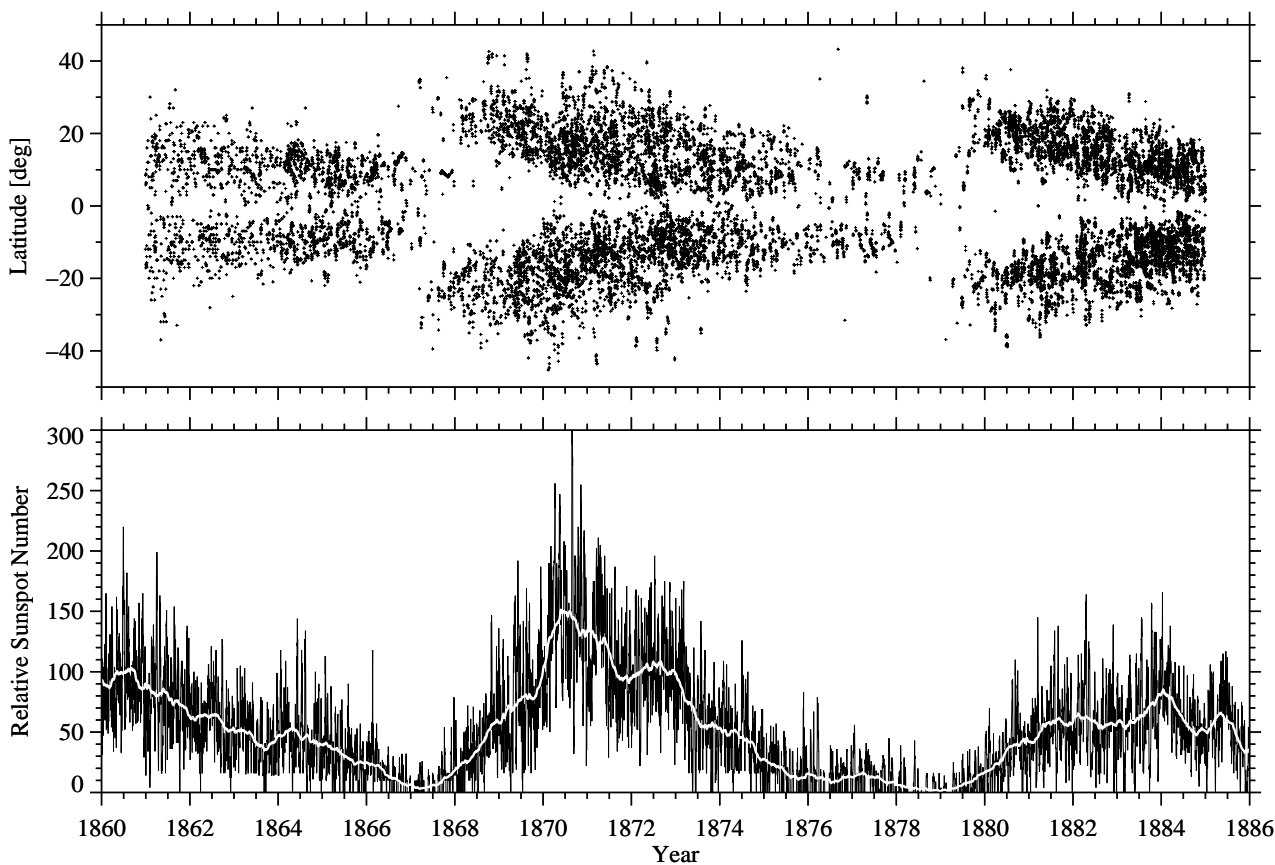}
\caption{The butterfly diagram (\textit{top}) contains the positions of all
sunspots from 1861 to 1884 listed in the electronic version of Sp\"orer's
tabulated measurements. The SIDC daily relative sunspot numbers from 1860 to
1885 (\textit{bottom}) are depicted for the daily values (\textit{black}) and a
200-day, sliding average (\textit{white}).}
\label{Fig:butterfly}
\end{figure*}

\subsection{Eliminating grid lines and identifying sunspots}  
\label{sec:eliminate}

The 3600$\times$800-pixel images contain grid lines, sunspots, and to some 
extent artifacts, i.e., dust particles, stains, or elongated fissures because of 
buckling pages. The goal is to identify the sunspots by removing  
artifacts and grid lines. First, the grayscale images are slightly smoothed and 
thresholded to yield binary masks. Small-scale artifacts with less than 
16~contiguous pixels are eliminated using standard tools for ``blob analysis'' 
\citep{Fanning2011}.

Second, removing grid lines is more complicated because they can overlap with 
sunspots. We start by removing the horizontal grid lines. The grid lines vary in 
width, but in general, they are not wider than seven pixels. Thus, we create a 
template, where the horizontal, seven-pixel-wide grid lines correspond to unity 
and the spaces in between to zero. We now exploit the fact that a sunspot 
superimposed on a grid line will broaden the line locally. We create a 
21$\times$21-pixel kernel, where the pixels of the vertical row through the 
center are set to \nicefrac{1}{21} and all other pixels to zero, so that the 
sum of all values is unity. The width of the kernel is chosen as three times 
the width of the grid lines. Convolving the binary mask with this kernel and 
thresholding the result leaves only areas where grid line intersect with 
sunspots. Simple logic operations are then used to eliminate the horizontal grid 
lines, and the procedure is repeated for the vertical grid lines. The algorithm 
sometimes leaves small artifacts of a few contiguous pixels, which are again 
removed with blob analysis tools.

The resulting binary mask contains only the sunspots, the numbers labeling the 
sunspot groups, and very few remaining artifacts. Until now, the whole procedure 
did not require any user interaction. In the final step, group labels and 
artifacts are removed by a point-and click routine. The result in 
Fig.~\ref{Fig:test2} shows the grayscale image multiplied by the binary mask, 
where ultimately only pores, sunspots, and active regions persist. Based on 
these results we create a butterfly diagram (Sect.~\ref{sec:butterfly}), and we 
discuss exemplarily how to calculate the tilt angles of the sunspot groups and 
 to determine other sunspot properties (Sect.~\ref{sec:position}). The 
log-normal probability density function (PDF) of the measured sunspots is 
compared in Sect.~\ref{sec:lognorm} with other frequency distributions taken 
from literature.

\begin{figure*}[t]
\centering
\includegraphics[width=\textwidth]{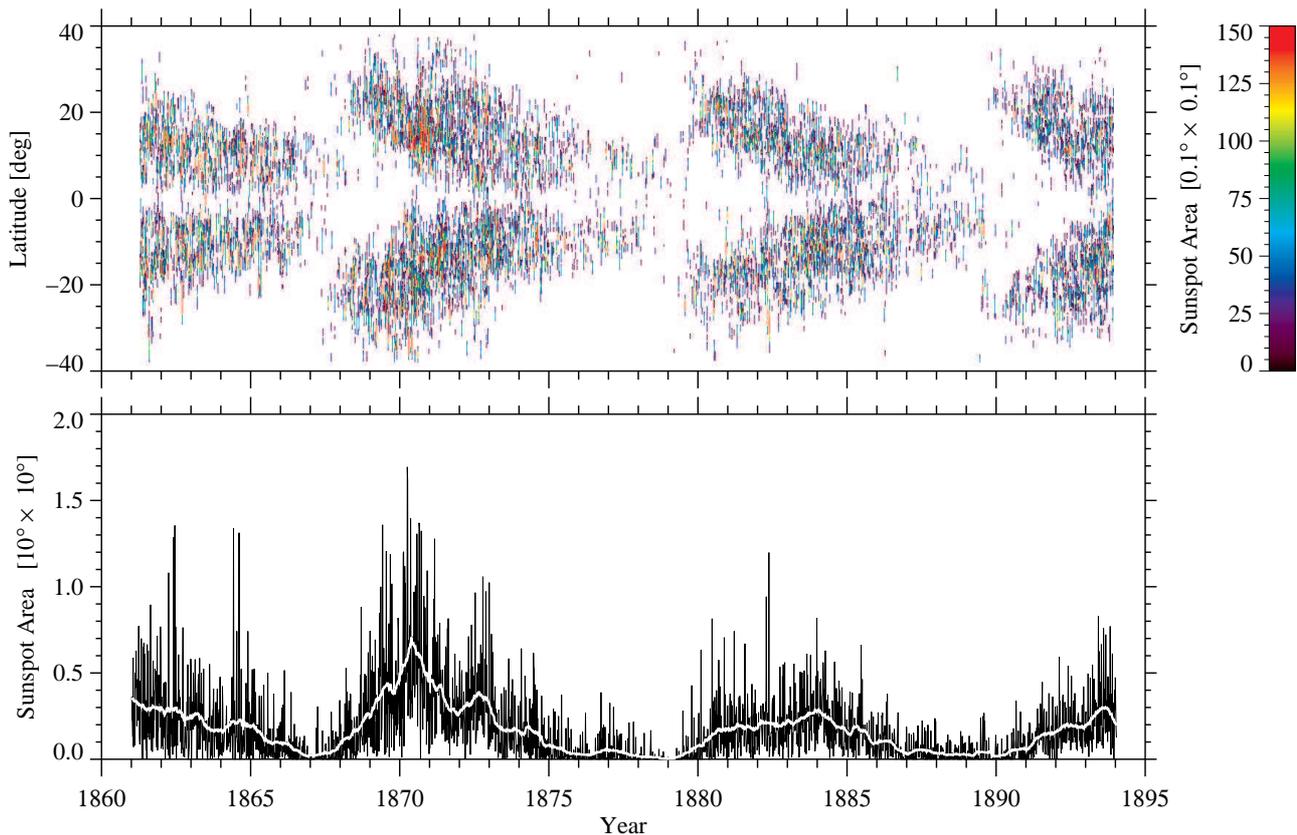}
\caption{Butterfly diagram (\textit{top}) with color-coded sunspot areas. The 
colors indicate the number of occurrence of sunspot pixels with a size of 
0.1$^{\circ} \times$ 0.1$^{\circ}$ within a 60$^{\circ}$-wide and 
0.4$^{\circ}$-high longitude-latitude bin. The sunspot area (\textit{bottom})   
is based on the number of pixels with a size of 0.1$^{\circ}\times0.1^{\circ}$ 
in a 60$^{\circ}$-wide longitude bin (\textit{black}). The white curve 
corresponds to a 200-day, sliding average to facilitate comparison with the 
lower panel of Fig.~\ref{Fig:butterfly}. The units on the ordinate combine 
10\,000 individual sunspot pixels.}   
\label{fig:fusion}
 \end{figure*}
 
\subsection{Butterfly diagram and relative sunspot number}\label{sec:butterfly}

We create a butterfly diagram (top panel of Fig.~\ref{Fig:butterfly}) from the
sunspot positions of Sp\"orer's measurements covering the period from 1861 to
1884. This diagram shows the latitudinal position of the sunspots in the
northern and southern hemispheres over a period of 24 years. In the beginning of
solar cycle No.~11 (1867--1878), new-cycle sunspots appear at high latitudes,
while old-cycle sunspots are still present at low latitudes, in particular in
the northern hemisphere. As the cycle progresses, more sunspot groups can be
observed at successively lower latitudes. In general, the distribution of
sunspots across latitude is more uniform in the southern hemisphere. The
sunspots' migration towards the solar equator over the solar cycle, where the
number of spots decreases, is today known as Sp\"orer's law.

The butterfly diagram is complemented by the daily relative sunspot numbers
(bottom panel of Fig.~\ref{Fig:butterfly}),  which are obtained from the Solar
Influences Data analysis Center 
(SIDC)\footnote{\href{http://sidc.oma.be/sunspot-data/dailyssn.php}
{www.sidc.oma.be/sunspot-data/dailyssn.php}}, Royal  Observatory of Belgium
\citep{Vanlommel2004, Clette2007}. The variation of the spot coverage with time
is much easier to visualize in a graph. In particular, solar cycle No.~11 shows
a pronounced double-peaked maximum \citep{Feminella1997} both in the daily
sunspot numbers and their 200-day average. Generally, the temporal evolution of
solar activity is well correlated in these two data sets.

The processed digital synoptic maps (Sect.~\ref{sec:eliminate}) offer another 
way to create a butterfly diagram based on the whole time series of the years 
1861 to 1894. In principle, one could concatenate all synoptic maps yielding a 
map with $(445\times3600)\times800 = 1\,602\,000 \times 800$ pixel (a 
1.3~gigapixel image!), which we transform into a super-synoptic map with 
$2670\times200$ pixels, i.e., each pixel contains the number of sunspots 
pixels in a 60$^\circ$-wide and 0.4$^\circ$-wide longitude-latitude bin. The 
size of a pixel is $0.1^\circ\times0.1^\circ$, thus, the frequency of occurrence 
of sunspot pixels corresponds to the sunspot area. The result of this 
compilation is shown in Fig.~\ref{fig:fusion}. Visually the butterfly diagrams 
of Figs.~\ref{Fig:butterfly} and \ref{fig:fusion} agree well. However, tight 
cluster of high sunspot counts are easier seen in Fig.~\ref{fig:fusion}, 
especially, in 1870 in the northern hemisphere. Projecting the two-dimensional 
frequency distribution onto the time axis produces a graph similar to the daily 
relative sunspot numbers in the bottom panel of Fig.~\ref{Fig:butterfly}. Note, 
however, that a 60$^\circ$-wide longitude bin matches about 4.5~days of a solar 
rotation period. This relationship is also used to compute a 200-day, sliding 
average, which again is closely correlated to the 200-day average of the SIDC 
daily relative sunspot numbers.

\subsection{Sunspot properties and active region tilt angles} 
\label{sec:position}

\begin{table*}[t]
\caption{Sunspot properties corresponding to Fig.~\ref{fig:spotgroup}.}
\begin{tabular}{ccccccccc}
	\hline\hline
Sunspot & Perimeter area & \rule{1mm}{0mm} & \multicolumn{2}{c}{Center 
coordinates} &  \rule{1mm}{0mm} & \multicolumn{3}{c}{Ellipse fitting} 
\rule{0mm}{3.5mm} \\
ID     &  $[1^\circ\times1^\circ]$ & &  $L$ & $b$ & & $a$ 
& $\varepsilon$ & $\theta$ \rule[-1.5mm]{0mm}{3mm}  \\
	\hline
$\phn 1$ & $8.05$ &  \rule[0mm]{0mm}{3.5mm} & $65.3\arcdeg$ &     
$-15.2\arcdeg$ &  & $2.08\arcdeg$ & $0.78$ & $-39.8\arcdeg$   \\
$\phn 2$ & $7.88$ &  & $72.7\arcdeg$ &     $-14.0\arcdeg$ &  & $1.73\arcdeg$ & 
$0.44$ & $-36.0\arcdeg$ \\
$\phn 3$ & $0.30$ &  & $69.4\arcdeg$ &     $-14.9\arcdeg$ &  & $0.36\arcdeg$ & 
$0.52$ & \phm $70.3\arcdeg$ \\
$\phn 4$ & $8.56$ &  & $67.5\arcdeg$ &     $-11.5\arcdeg$ &  & $2.34\arcdeg$ & 
$0.74$ & $-80.7\arcdeg $\\
$\phn 5$ & $0.28$ &  & $69.3\arcdeg$ &     $-12.9\arcdeg$ &  & $0.36\arcdeg$ & 
$0.61$ & \phm $77.9\arcdeg$ \\
$\phn 6$ & $0.22$ &  & $75.7\arcdeg$ &     $-12.6\arcdeg$ &  & $0.35\arcdeg$ & 
$0.72$ & $-51.4\arcdeg$ \\
$\phn 7$ & $1.23$ &  & $76.7\arcdeg$ &     $-11.2\arcdeg$ &  & $1.17\arcdeg$ & 
$0.90$ & \phm $79.4\arcdeg$ \\
$\phn 8$ & $7.70$ &  & $80.1\arcdeg$ &     $-10.8\arcdeg$ &  & $1.86\arcdeg$ & 
$0.62$ & $-23.8\arcdeg$ \\
$\phn 9$ & $0.45$ &  & $72.1\arcdeg$ &     $-11.3\arcdeg$ &  & $0.41\arcdeg$ & 
$0.28$ & $-34.9\arcdeg$ \\
    $10$ & $0.82$ & \rule[-1.5mm]{0mm}{3mm} & $77.9\arcdeg$ & \phn 
$-9.7\arcdeg$ &  & $0.82\arcdeg$ & $0.90$ & $-59.2\arcdeg$
\\
\hline
\end{tabular}
\label{tab:02}
\end{table*}

\begin{figure}[t]
\centering
\includegraphics[width=\columnwidth]{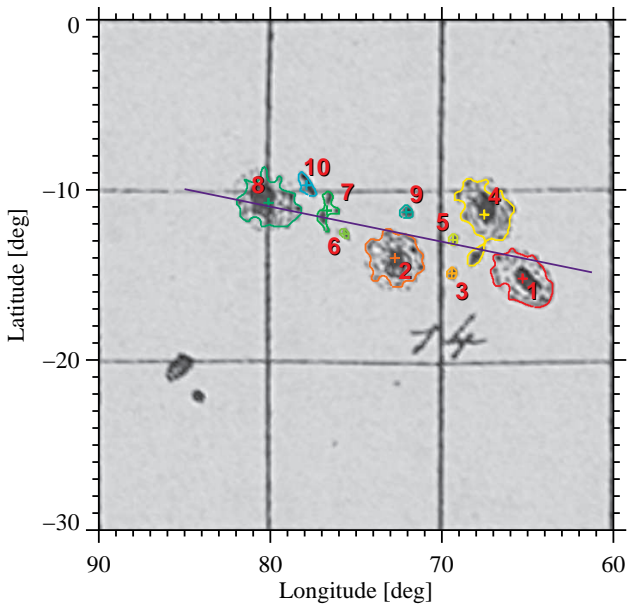}
\caption{Sunspot group No.~14 with ten identified sunspots. Their properties are 
given in Tab.~\ref{tab:02}. A linear fit (\textit{blue line}) to all pixels 
belonging to the sunspot group indicates the region's tilt angle.} 
\label{fig:spotgroup}
 \end{figure}

Binary masks like the one used to create Fig.~\ref{Fig:test2} are the starting 
point for determining sunspot properties. Application of blob analysis tools 
\citep{Fanning2011} returns parameters such as the sunspot area and the 
perimeter in pixels, the coordinates of the sunspot's center-of-gravity (CoG), 
and from ellipse fitting the semi-major axes $a$, the numerical eccentricity 
$\varepsilon$, and the orientation $\theta$ of an ellipse encompassing the 
sunspot. The results are visualized for sunspot group No.~14 in 
Fig.~\ref{fig:spotgroup}, and the quantitative results are presented in 
Tab.~\ref{tab:02}. Still some problems remain. Not all sunspots are recognized 
as single spots. Nearby spots are often identified as one single spot. As a 
consequence, the Wolf number will be smaller than expected. In the example, the 
correct Wolf number is 24, but the Wolf number based on our algorithm is only 
20.

Furthermore, we measure the tilt angle of the group, which is $11.6\arcdeg$, by 
fitting a line to all pixels belonging to the spot group. Joy's law 
\citep{McClintock2013} indicates that a sunspot group will establish a tilt 
angle with the leading spot closest to the equator. In this example, the leading 
spot has the largest distance to the equator. However, the spot group is complex 
and contains four mature sunspots. Without any additional information about 
the group's magnetic configuration or temporal evolution, it is impossible to 
decide, if the group emerged as one or represents the superposition of two 
bipolar regions.

Visual inspection of the synoptic maps gives the impression that regular 
sunspots with well established penumbrae are drawn too large. This might also 
apply to pores, where pencil marks might already be too coarse to correctly 
indicate the actual size. The positions, however, of spots are accurately 
represented in the synoptic maps, with the exception of the errors already noted 
in Sect.~\ref{sec:errors}.

\subsection{Probability density function of sunspot areas} \label{sec:lognorm}

Size and location of individual sunspots were already used in creating the 
butterfly diagram in Fig.~\ref{fig:fusion}. Morphological image 
processing  additionally provides access to another important characteristics 
of sunspots, i.e., the PDF of their areas over the course of 33~years of 
observations by Sp\"orer.

A pixel with a size of 0.1\arcdeg$\times$0.1\arcdeg\ corresponds at the 
intersection of equator and central meridian to an area of about 1.5~Mm$^2$ or 
0.5~$\mu$HS (one millionth of the visible solar surface). Lacking a clear 
description of how the sunspot drawings were transferred to an equidistant 
longitude-latitude grid, it remains unclear, if geometrical projection effects 
were taken properly into account. However, projection effects can only introduce 
an error of about 25\%. The annual variations related to Earth's orbit 
around the Sun and the solar $B_0$-angle (heliographic latitude of the central 
point of the solar disk) is only about 0.6\%, which is negligible compared with 
other systematic errors. For example, thresholding and morphological 
dilation/erosion operations can lead to erroneous sunspot boundaries, thus, 
overestimating sunspot areas by at most a few tens of percent. Even 
taken together, these effects cannot explain the apparently oversized sunspots 
in the synoptic maps. Therefore, we decided to compare our PDF of sunspot areas 
with previous studies.

\begin{SCfigure*}
\includegraphics[width=0.74\textwidth]{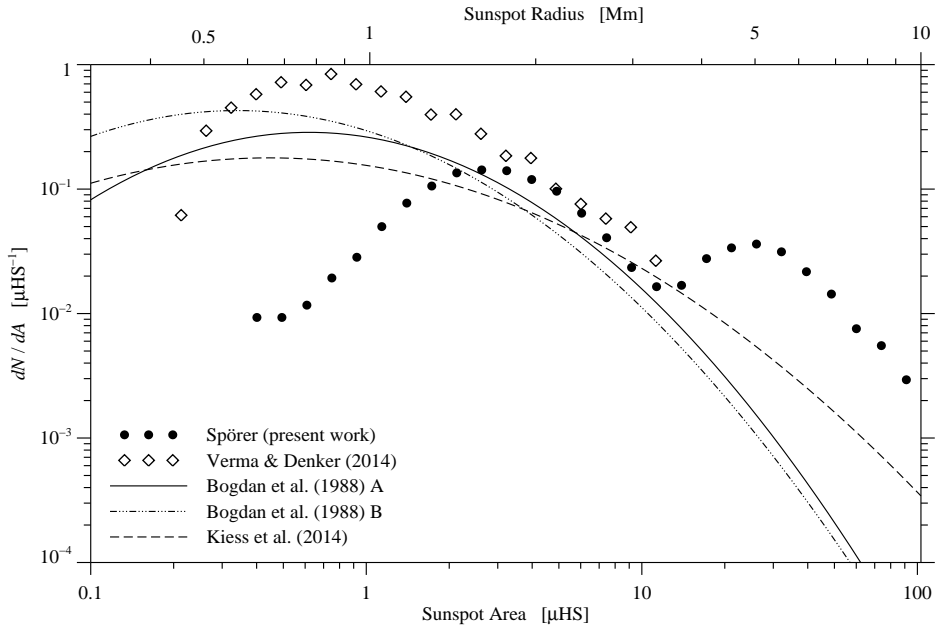}
\caption{Probability density functions of areas and radii for sunspots and 
pores (\textit{symbols}) and their comparison to previous studies 
(\textit{curves}).}
\label{Fig:lognorm}
\end{SCfigure*}

Log-normal PDFs of pores and umbral areas are described in \citet{Bogdan1988} 
and \citet{Kiess2014} for Mount Wilson data (1917--1982) and contemporary 
observations from space (Solar Dynamics Observatory 
\citep[SDO,][]{Pesnell2012}, May 2010 to October 2012), respectively. Their PDFs 
are presented in Fig.~\ref{Fig:lognorm} as continuous curves, where we used the 
notation and normalization scheme of the PDF given by \citet{Kiess2014}
\begin{align}
\ln\left(\frac{dN}{dA}\right) = - \frac{(\ln A - \ln\langle A \rangle)^2}{2 
\ln \sigma_A} + \ln\left(\frac{dN}{dA}\right)_{\text{max}} \label{eq:lognorm},
\end{align}
where $A$ is the umbral/pore size, $(dN/dA)$ is the density function,  $\langle 
A \rangle$ is the mean, and $\sigma_A$ is the width of the distribution. A 
distinguishing characteristics of these PDFs is the spacial resolution, i.e., 
the minimum spot size of $2$ and $0.5~\mu$HS, respectively. Using high 
resolution G-Band images of the Japanese \textit{Hinode} mission, 
\citet{Verma2014} derived the statistical properties of pores for the time 
period of October 2006 to August 2013 (open diamonds in Fig.~\ref{Fig:lognorm}). 
Here, the minimum spot size is just $0.25~\mu$HS, which is a good lower 
threshold for magnetic field concentrations still capable of inhabiting the 
convective energy transport.

Matching Sp\"orer's sunspot data to the aforementioned PDFs requires an 
appropriate scaling of the sunspot areas. The PDF based on Sp\"orer's 
data is double-peaked because the morphological image processing 
algorithm detects both pores (filled pencil marks) and sunspots (filled umbrae 
and dotted penumbrae surrounded by a thin boundary). The first peak corresponds 
to the pores and the second peak to the sunspots. We use the minimum between the 
two peaks as a threshold, which yields an average spot area of  38.9~$\mu$HS 
for spots with an area smaller than $100~\mu$HS. The average size of pores and 
umbrae in \citet{Kiess2014} is 10.7~$\mu$HS. Therefore, we conclude that the 
ratio of 13.3 is a conservative estimate of the area scaling factor 
between SDO and Sp\"orer data.

Using this scaling parameter, we place the Sp\"orer data (solid circles) in the 
log-log frequency distribution shown in Fig.~\ref{Fig:lognorm}. With this 
correction, the frequencies of occurrence between 2 and 10~$\mu$HS are very 
similar to the aforementioned studies. Above 10~$\mu$HS only sunspots with 
umbrae and penumbrae are counted, thus, the curve is shifted parallel to the 
right. The strong deviation to lower frequency of the Sp\"orer PDF below 
$2~\mu$HS can be attributed to the limited spacial resolution offered by 
Sp\"orer's telescopes. Under ideal conditions the diffraction-limited  
resolution is about 1000~km on the solar surface. 
Therefore, a large fraction of small sunspots and pores will be missed in 
Sp\"orer's observations. Despite the limitations imposed by the scaling factor, 
we are in this manner able to present a self-consistent approach to place   
Sp\"orers historic sunspot data into the parameter space of studies from the 
$20^{\mathrm{th}}$ and $21^{\mathrm{st}}$ century.

%
%

\section{Summary}

There are several ways of extracting information about position and size from 
historical sunspot drawings. A fully manual method was employed by 
\citet{Arlt2009a} and \citet{Arlt2013} on the drawings by Staudacher and 
Schwabe, respectively. A semi-automatic procedure was used by 
\citet{Cristo2011}, who analyzed the sunspot observations by Ludovic Zucconi in 
1754--1760. They picked spots manually and determined their area automatically. 
In the present work, we put forward a procedure for automatic extraction of 
positions and sizes of sunspots in historical records and applied it to 
observations from 1861--1894 by Gustav Sp\"orer. It was possible to process the 
images of his synoptic maps and to obtain good results -- despite some 
difficulties related to the scale of the sunspot in the drawings. Ultimately, a 
150-year-old data set of sunspot observations is now available for contemporary 
data analysis facilitating the study of the Sun's activity in the past. The data 
presented in this article are available upon request by contacting one of the 
authors.

%
%

\acknowledgements CD was supported by grant DE~787/3-1 of the Deutsche
Forschungsgemeinschaft (DFG). The authors thank Regina von Berlepsch for her
support in the library of the Leibniz-Institut f\"ur Astrophysik Potsdam.
We also express our gratitude to Drs.\ Horst Balthasar and Jan Ryb\'ak for 
carefully reading the manuscript and for comments improving this work.

%
%


\begin{thebibliography}{36}
\expandafter\ifx\csname natexlab\endcsname\relax\def\natexlab#1{#1}\fi

\bibitem[\protect\citeauthoryear{Arlt}{2008}]{Arlt2008}
    Arlt, R. 2008, Sol. Phys., 247, 399
\bibitem[\protect\citeauthoryear{Arlt}{2009a}]{Arlt2009a}
    Arlt, R. 2009a, Sol. Phys., 255, 143
\bibitem[\protect\citeauthoryear{Arlt}{2009b}]{Arlt2009b}
    Arlt, R. 2009b, AN, 330, 311
\bibitem[\protect\citeauthoryear{Arlt}{2011}]{Arlt2011a}
    Arlt, R. 2011, AN, 332, 805
\bibitem[\protect\citeauthoryear{Arlt et al.}{2013}]{Arlt2013}
    Arlt, R., Leussu, R., Giese, N., et al. 2013, Mon. Not. R. Astron. Soc., 
    433, 3165
\bibitem[\protect\citeauthoryear{Bard et al.}{1997}]{Bard1997}
Bard, E., Raisbeck, G.~M., Yiou, F., \& Jouzel, J. 1997, Earth Plan.
Sci. Lett., 150, 453
\bibitem[\protect\citeauthoryear{Balthasar \& Fangmeier}{1988}]{Balthasar1988}
    Balthasar, H., \& Fangmeier, E. 1988, 
    Astron. Astroph., 203, 381
\bibitem[\protect\citeauthoryear{Baumann \& Solanki}{2005}]{Baumann2005}
    Baumann, I., \& Solanki, S.~K. 2005, 
    Astron. Astroph., 443, 1061 
\bibitem[\protect\citeauthoryear{Bogdan et al.}{1988}]{Bogdan1988}
    Bogdan, T.~J., Gilman , P.~A.,  Lerche , I., \& Howard, R. 1988, 
    ApJ, 327, 451 
\bibitem[\protect\citeauthoryear{Carrington}{1863}]{Carrington1863}
    Carrington, R.~C. 1863, Observations of the Spots on the Sun: from
      November 9, 1853, to March 24, 1861, Made at Redhill
    (Williams and Norgate, London)
\bibitem[\protect\citeauthoryear{Clette et al.}{2007}]{Clette2007}
    Clette, F., Berghmans, D., Vanlommel, P., van der Linden, R.~A.~M.,
	 Koeckelenbergh, A., \& Wauters, L. 2007, Adv. Space Res., 40, 919
\bibitem[\protect\citeauthoryear{Cristo et al.}{2011}]{Cristo2011}
    Cristo, A., Vaquero, J.~M., \& S\'anchez-Bajo, F. 2011, J. Atm. Sol.-Terr. 
Phys., 73, 187
\bibitem[\protect\citeauthoryear{Eddy}{1976}]{Eddy1976}
    Eddy, J.~A. 1976, Science, 192, 1189 
\bibitem[\protect\citeauthoryear{Fanning}{2011}]{Fanning2011}
    Fanning, D.~W. 2011, Coyote's Guide to Traditional IDL Graphics
    (Coyote Book Publ., Fort Collins, Colorado)
\bibitem[\protect\citeauthoryear{Feminella \& Storini}{1997}]{Feminella1997}
    Feminella, F., \& Storini, M. 1997, Astron. Astroph., 322, 311 
\bibitem[\protect\citeauthoryear{Gonzalez \& Woods}{2002}]{Gonzalez2002}
    Gonzalez, R.~C., \& Woods, R.~E. 2002, Digital Image Processing
    (Prentice-Hall, Upper Saddle River, New Jersey)
\bibitem[\protect\citeauthoryear{Kiess et al.}{2014}]{Kiess2014}
    Kiess, C., Rezaei, R., \& Schmidt, W. 2014, Astron. Astroph, 565, A52
\bibitem[\protect\citeauthoryear{Lee \& Schachter}{1980}]{Lee1980}
    Lee, D.~T., \& Schachter, B.~J. 1980, J.\ Int.\ Comp.\ Inform.\ Sci., 9, 219
\bibitem[\protect\citeauthoryear{Lee}{1986}]{Lee1986}
    Lee, J.~S. 1986, Opt.\ Eng., 25, 636
\bibitem[\protect\citeauthoryear{McClintock \& Norton}{2013}]{McClintock2013}
McClintock, B.~H., \& Norton, A.~A. 2013, Sol. Phys., 287, 215
\bibitem[\protect\citeauthoryear{Noback \& Noback}{1851}]{Noback1851}
  Noback, C., \& Noback, F., 1851, Vollst\"andiges Taschenbuch der M\"unz-,
  Maa\ss- und Gewichts-Verh\"altnisse,
  der Staatspapiere, des Wechsels- und Bankwesens und der Usanzen aller L\"ander
  und Handelspl\"atze (Brockhaus, Leipzig)
\bibitem[\protect\citeauthoryear{Park \& Schowengerdt}{1983}]{Park1983}
    Park, S.~K., \& Schowengerdt, R.~A. 1983, Comp.\ Vis.\ Graph.\ Image Proc., 
    23, 258
\bibitem[\protect\citeauthoryear{Pesnell et al.}{2012}]{Pesnell2012}
    Pesnell, W.~D., Thompson, B.~J., \& Chamberlin, P.~C. 2012, Sol. Phys., 
    275, 3
\bibitem[\protect\citeauthoryear{Sp\"orer}{1861}]{Spoerer1861}
    Sp\"orer, G. 1861, Beobachtungen von Sonnenflecken und daraus
    abgeleitete Elemente der Rotation der Sonne (Dietze, Anclam, Germany)
\bibitem[\protect\citeauthoryear{Sp\"orer}{1874}]{Spoerer1874}
    Sp\"orer, G. 1874, Pub.\ Astron.\ Ges., 13
\bibitem[\protect\citeauthoryear{Sp\"orer}{1878}]{Spoerer1878}
    Sp\"orer, G. 1878, Pub.\ Astrophys.\ Obs.\ Potsdam, 1
\bibitem[\protect\citeauthoryear{Sp\"orer}{1880}]{Spoerer1880}
    Sp\"orer, G. 1880, Pub.\ Astrophys.\ Obs.\ Potsdam, 5
\bibitem[\protect\citeauthoryear{Sp\"orer}{1886}]{Spoerer1886}
    Sp\"orer, G. 1886, Pub.\ Astrophys.\ Obs.\ Potsdam, 17, 220
\bibitem[\protect\citeauthoryear{Sp\"orer}{1887}]{Spoerer1887}
    Sp\"orer, G. 1887, Vierteljahresschr.\ Astron.\ Ges., 22, 323
\bibitem[\protect\citeauthoryear{Sp\"orer}{1894}]{Spoerer1894}
    Sp\"orer, G. 1894, Pub.\ Astrophys.\ Obs.\ Potsdam, 32
\bibitem[\protect\citeauthoryear{Usoskin et al.}{2007}]{Usoskin2007}
    Usoskin, I.~G., Solanki, S.~K., \& Kovaltsov, G.~A. 2007, Astron. Astroph., 
    471, 301
\bibitem[\protect\citeauthoryear{W\"ohl \& Balthasar}{1989}]{Woehl1989}
    W\"ohl, H., \& Balthasar, H. 1989, 
    Astron. Astroph., 219, 313
\bibitem[\protect\citeauthoryear{Wolf}{1861}]{Wolf1861}
    Wolf, R. 1861, Astron. Mitt. Eidgen. Sternw. Z\"urich, 2, 83
\bibitem[\protect\citeauthoryear{Vanlommel et al.}{2004}]{Vanlommel2004}
    Vanlommel, P., Cugnon, P., van der Linden, R.~A.~M., Berghmans, D., 
    \& {Clette}, F. 2004, Sol. Phys., 224, 113
\bibitem[\protect\citeauthoryear{Verma \& Denker}{2014}]{Verma2014}
    Verma, M., \& Denker, C. 2014, Astron. Astroph., 563, A112 
\bibitem[\protect\citeauthoryear{Vogel}{1895}]{Vogel1895}
    Vogel, H. 1895, ApJ, 2, 239
\end{thebibliography}
\end{document}